\documentstyle[12pt]{article}
\newcommand{\tilda}{\tilde}
\newcommand{\bra}{\langle}
\newcommand{\ket}{\rangle}
\newcommand{\eq}{\begin{equation}}
\newcommand{\en}{\end{equation}}
\newcommand{\eqn}{\begin{eqnarray}}
\newcommand{\enn}{\end{eqnarray}}
\newcommand{\lm}{\lambda}
\newcommand{\pa}{\partial}
\newcommand{\A}{\alpha}

\begin{document}

\begin{titlepage}
\null

\begin{flushright}
hep-th/9611097   \\
UT-Komaba/96-25
\end{flushright}
\vspace{0.5cm} 
\begin{center}
{\Large \bf 
SUSY $N=2$ hyperelliptic curve from 
 
$N=1$ effective potential 

\par}
\lineskip .75em
\vskip2.5cm
\normalsize
{\large Takuhiro Kitao} 
\vskip 1.5em
{\it Institute of Physics, University of Tokyo, Komaba, Meguro-ku, Tokyo 153, Japan}\\
E-mail address: kitao@hep1.c.u-tokyo.ac.jp
\vskip3cm
{\bf Abstract}
\end{center} \par
We derive the singularity conditions of the $N=1$ generalized (general yukawa couplings and quark masses) form of hyperelliptic curves of $SU(N_{c})$ with $N_{f}$ flavors. 
The results reproduce the known form of $N=2$ curves when the yukawa couplings and the quark masses reduce to those of $N=2$. 
We obtained these curves by determining the dependence of the unbroken $SU(2)$ gaugino condensation on the couplings in the moduli source terms which break $N=2$ SQCD to $N=1$ $SU(N_c)$ gauge theory with the quarks and the adjoint matter, $\Phi$. 
The degenerate component of the diagonalized classical vacuum expectation value of $\Phi$ is shown to be explicitly written in terms of these couplings, which enables us to determine the form of the gaugino condensation.

\end{titlepage}
\renewcommand{\thefootnote}{\arabic{footnote}}
\setcounter{footnote}{0}
\baselineskip=0.7cm

The recent developments of $N=2$ SUSY gauge theory enable us to obtain exact descriptions of the low energy strong coupling region. According to them, the theories are parametrized by the vacuum expectation values of the gauge invariant operators which form the quantum moduli space. The singularities of the quantum moduli space correspond to the appearance of massless solitons. Therefore when massless solitons appear, the (hyperelliptic) curves which describe the quantum moduli space have vanishing cycles. If we add the moduli source terms which break $N=2$ to $N=1$, all the $N=2$ vacua without massless solitons lift and only the $N=2$ vacua with massless solitons remain as $N=1$ vacua \cite{Wit}, \cite{aps}.

On the other hand, we may start from the $N=1$ theory with the superpotential containing the $N=2$ tree-level superpotential and the moduli source terms which reduce $N=2$ SUSY to $N=1$. When we investigate the low energy strong coupling region of the theory, we can make use of the technique to decide the form of $N=1$ effective superpotentials by symmetry. From the low energy effective superpotential determined in this way, we can derive the moduli, namely, the v.e.v.s of the gauge invariant operators which describe the quantum vacua. We can show that a hyperelliptic curve becomes singular at these v.e.v.s and the curve corresponds with that of $N=2$ theory in the $N=2$ limits of the couplings.

Indeed, for $SU(N_c)$ pure super Yang-Mills, Elitzur et al. \cite{Eli} have obtained the $N=1$ effective superpotential from which the $N=2$ curve, \cite{Kle},  \cite{Far} is reproduced.
The approach has been generalized to the other classical groups in the case of pure SYM in ref \cite{ter}. Consistency with $N=2$ results has been further checked for $SO(5)$ with one flavor in \cite{Gid} and the authors have also applied this method to $G_{2}$ SUSY gauge theory. 
  
 Our purpose in this paper is to examine the case of SQCD ($SU(N_c)$)
with general flavors. The authors of \cite{Eli} have also tried this
case. They ``integrate in'' \cite{Int} the adjoint matter from the $N=1$ low energy $SU(2)$ SQCD effective potential and get the superpotential with the
adjoint matter in the high energy. This potential leads to $SU(2)$
SQCD effective potential in the low energy when the adjoint matter is
integrated out around the classical vacua with $SU(2) \times
U(1)^{N_c-2}$ gauge symmetry. Using the equations of motion for this
potential, they reproduced the $N=2$ hyperelliptic curve \cite{Arg} 
in the case of one flavor SQCD. But the known $N=2$ curves for SQCD 
with general flavors have not been re-derived by this method. 
We shall derive the
singularity conditions for the curves of SQCD from the low energy
superpotential which is obtained by integrating all the matter. This
low energy potential contains the $SU(2)$ gaugino condensation. The
key point is that we can determine the dependence of this gaugino
condensation on the couplings $g_r$ ($r=2,3,\cdots N_c$) in Eq.(\ref{$1$})
in this paper. As the result of that , we can show that the method
which the authors of \cite{Eli} have used for SYM successfully can
also be applicable to the case of SQCD with general flavors. These
curves derived by this method reproduce the curves of $N=2$ SQCD when
the couplings reduce those of $N=2$ SQCD.
   
 We first briefly summarize the methods of our derivation. In this paper we denote quarks by $Q^r$ and anti-quarks by $\tilde Q^s$ which are the $N=1$ chiral superfields in the fundamental and anti-fundamental representations of $SU(N_c)$ gauge symmetry, where $r, s=1,2,\cdots N_f$ are flavor indices. They make up $N=2$ hypermultiplets together. We also denote by $\Phi$ an $N=1$ adjoint chiral superfield which is in an $N=2$ vector multiplet.  
\vskip 0.8em
$1\ \ $ By starting from the following superpotential,
\begin{equation}
W_{tree}=\sum _{k=2}^{N_c}\frac{g_{k}{\rm Tr}\Phi^{k}}{k} +\lambda Q\Phi \tilde Q+m_{Q}Q \tilda Q,
\label{$1$}
\end{equation}
we can perturb $N=2$ SQCD into $N=1$ theory. Here, $\lambda$, $m_{Q}$ are the couplings with the flavor indices, \footnote{We abbreviate flavor indices of $\lambda$ and $m_{Q}$, and the summation over them in (\ref{$1$}).} and $g_{k}$ is the coupling to the gauge invariant operator, $\frac{{\rm Tr}\Phi^{k}}{k}$. Then assuming the classical vacua, 
$\bra \tilda {Q} \ket =0$, $\bra Q \ket =0$
for all the quarks, we get the classical v.e.v of $\Phi$ , $\Phi_{cl}$ from the equations of motion . Classically, the generic unbroken nonabelian gauge symmetry is $SU(2)$ and we can determine the degenerate component $M$ of $\Phi_{cl}={\rm diag}(M,M,M_3,M_4,\cdots M_{N_c})$, where $M \neq M_a \neq M_b$, ($a \neq b$) from the equation of motion.  
\vskip 0.8em
$2\ \ $ Next, we integrate out massive particles, except the quarks which transform as the fundamental and anti-fundamental representations of $SU(2)$ (we denote these quarks as the $SU(2)$ quarks below.) around the chosen vacua  and add the low energy effective superpotential of $N=1$ $SU(2)$ massless SQCD, $W_{d}$. In addition to them, we must consider the other quantum corrections, $W_\Delta$ using the notation of ref \cite{Int}.  $W_\Delta$ is the potential which may generate in the low energy theory because of the quantum effects of the high energy theory. We can not exclude this term by the symmetry and the classical limits. We assume below that $W_{\Delta}$ is zero and compare the results under this assumption with the exact results. 
Finally, we get the low energy superpotential with only the $SU(2)$ quarks, $W_{L} =W_{cl}+W_{d}$.
Here, we denote $W_{cl}$ as the potential substituted in $W_{tree}$ by $\Phi_{cl}$ and the quark v.e.v.s except the $SU(2)$ quarks. Here, we assume that the masses of the quarks except the $SU(2)$ quarks , that is $M_a$ ($a=3,4,\cdots N_c$), are very large, so that the v.e.v.s of these quarks are zero. 
\vskip 0.8em
$3\ \ $ Then we integrate out the $SU(2)$ quarks and get the low energy superpotential $W_{LL}$ which contains the $SU(2)$ gaugino condensation. We can determine this gaugino condensation in terms of $g_{N_c}$, $g_{N_c-1}$, $m_Q$ and the dynamical scale of the original theory, $\Lambda$. Here we have to limit the range of parameters so that the assumption in 1, $\bra \tilda {Q} \ket =0$, $\bra Q \ket =0$ is compatible with non zero v.e.v.s of the $SU(2)$ invariants  which consist of the $SU(2)$ quarks. Moreover in the case of $N_f>3$, these v.e.v.s of the $SU(2)$ invariants  must be far away from some special points, where massless solitons ( dual quarks ) contribute. But the vacua we analyze are most generic in the allowed range, so the results become consistent with $N=2$ theory and may be exact even in the range outside the above restriction because of holomorphy. 
\vskip 0.8em
$4\ \ $ By taking the derivative of the $W_{LL}$ with respect to $g_{r}$, we obtain the quantum moduli as in the case of SYM in \cite{Eli}. In this method, they are only the semi-classical moduli, but in fact it will turn out to be exact. In this way we can relate the classical moduli to the quantum moduli.
\vskip 0.8em
$5\ \ $ We can then write the characteristic equation for $\Phi_{cl}$ in terms of the quantum moduli, which leads to the condition of vanishing cycles of $N=2$ curve.
\vskip 0.8em
 Now let us proceed to calculations according to the above scenario. 
The classical equation of motion for $\Phi$ is 
\begin{equation}
\sum_{k=2}^{N_c} g_{k}\Phi^{k-1}-\frac{1}{N_{c}}\sum_{k=2}^{N_c}g_{k} {\rm Tr}\Phi^{k-1}=0.
\label{$4$}
\end{equation}
The second term has its origin in the fact that $\Phi$ is traceless. Namely, the equation of motion for $\Phi$ needs a Lagrange multiplier to take account into this constraint. The equation (\ref{$4$}) is the form after the Lagrange multiplier is eliminated. We define an equation by
\begin{equation}
\sum_{k=2}^{N_c} g_{k}x^{k-1}-\frac{1}{N_{c}}\sum_{k=2}^{N_c}g_{k} {\rm Tr}\Phi_{cl}^{k-1}=0.
\label{$5$}
\end{equation}
Below we often define $u_{k}=\bra \frac{{\rm Tr}\Phi^{k}}{k} \ket$, and $u^{cl}_{k}=\frac{{\rm Tr}\Phi_{cl}^{k}}{k}$. Let us consider the most generic case in which the $N_c-1$ components of the diagonalized classical solution of (\ref{$4$}), $\Phi_{cl}={\rm diag}(M_1,M_2,\cdots M_{N_c})$  are different from each other, say $M_a \neq M_b$ ($a \neq b$, $a,b=2,3,4\cdots N_c$). Then (\ref{$5$}) has $N_c-1$ solutions which can be identified with $N_c-1$ different components of the diagonalized classical solution of (\ref{$4$}). Therefore the diagonalized solution of (\ref{$4$}) has two components with the same value, $M$, that is $\Phi_{cl}={\rm diag}(M,M,M_3,\cdots M_{N_c})$, where $M \neq M_a$, $M_a \neq M_b$, ($a\neq b$, $a,b=3,4,\cdots N_c$).  The same logic as this is also applicable to other gauge groups besides $SU(N_c)$ and there is necessarily unbroken nonabelian gauge symmetry in any classical vacua \cite{ter}. 

In the particular case of $SU(N_c)$, fortunately, any component can be explicitly represented as minus the sum of the other $N_{c}-1$ solutions because the trace of $\Phi_{cl}$ is zero. So we can see from (\ref{$5$}) that the sum of the $N_c-1$ different components is $-\frac{g_{N_c-1}}{g_{N_c}}$. Thus we have shown that there are always two $\frac{g_{N_c-1}}{g_{N_c}}$s among the $N_c$ components of the diagonalized classical solution of (\ref{$4$}) when classically  $SU(2) \times U(1)^{N_c-2}$ is unbroken. We have assumed here the v.e.v.s of quarks are zero. Below we define $\frac{g_{N_c-1}}{g_{N_c}}$ as $M$. 

As a first example, let us consider the case of one flavor. After decomposing
$\Phi=\Phi_{cl}+\delta\Phi$, we substitute this into (\ref{$1$}),\footnote{In (\ref{delta}) we consider the potential that the Lagrange multiplier is not eliminated.}
\eq 
\sum_{k=1}^{N_c}\frac{ g_{k}{\rm Tr}{\Phi^{k}}}{k}=\sum_{k=1}^{N_c}\frac{ g_{k}{\rm Tr}{\Phi^{k}}_{cl}}{k} + \frac{1}{2} \sum_{k=1}^{N_c} g_{k}{\rm Tr}\Big(\delta\Phi^2 \Phi^{k-2}_{cl}\Big)(k-1)
   +  O(\delta\Phi^3).\label{delta}
\en
In particular, the mass term for the fluctuation along the unbroken $SU(2)$ is \cite{ter}, \cite{Kut}
\eqn
\frac{1}{2}  {\it m_{{\rm SU(2)}}} {\rm Tr}\delta\Phi^2_{{\rm SU(2)}}
    &=& \frac{1}{2} \sum_{k=1}^{N_c} (k-1)g_{k}{\rm Tr} \Big(\delta\Phi^2_{{\rm SU(2)}} \Phi^{k-2}_{cl}\Big)
    = \frac{1}{2} \sum_{k=1}^{N_c} (k-1)g_{k} {\rm Tr}\Big(\delta\Phi^2_{{\rm SU(2)}}\Big) M^{k-2}
    \nonumber \\
&=& \frac{1}{2} g_{N_c}\Big(\prod_{a=3}^{N_c} (M-M_a)\Big) {\rm Tr}\delta\Phi^2_{{\rm SU(2)}}. \nonumber
\enn
We assume that ${\it m_{{\rm SU(2)}}}$ is large compared with the
dynamical scale of the original theory, $\Lambda$. The quarks except
the $SU(2)$ quark have masses of order, $M_a$ ($a=3,4,\cdots N_c$) and we
also assume $M_a \gg \Lambda$.  Then we integrate out $\delta\Phi$ and
these quarks, assuming that ${\it m_{{\rm
SU(2)}}}$ and $M_a$ are so large that we can ignore the interaction
terms of $\delta \Phi$ and all the quarks, $\lambda Q \delta \Phi \tilde Q$.
 We denote the dynamical
scale of the low energy $SU(2)$ with one flavor, $\Lambda_{L}$ by the
relation, $\Lambda_{L}^{5}=\Lambda^{2N_c-1}{g_{N_c}}^{2}$ \cite{ter}
\cite{Kut}.  Finally we obtain the superpotential with only the
$SU(2)$ quark,
\begin{equation}
W_{L}=(\lambda M +m_{Q})Q^{\A}\tilda{Q}_{\A} + \frac{\Lambda_{L}^{5}}{Q^{\A} \tilda{Q}_{\A}
}+\sum_{k=2}^{N_c}\frac{ g_{k}{\rm Tr}{\Phi^{k}}_{cl}}{k}+W_{\Delta},
\label{$8$}
\end{equation}
where $\A=1,2$ mean the $SU(2)$ indices. The first term originates by substituting $\Phi_{cl}$ into $W_{tree}$ and the second is the $SU(2)$ dynamically generated term. In this step, we assume $W_{\Delta}$ is zero under the above assumption, $M_a$, ${\it m_{{\rm SU(2)}}}$ $\gg$ $\Lambda$ as in \cite{Eli}. Then, we integrate the $SU(2)$ quark, and we obtain the effective potential with the $SU(2)$ gaugino condensation,
\begin{equation}
W_{LL}=\sum_{k=2}^{N_c}\frac{g_{k}{\rm Tr}{\Phi^{k}_{cl}}}{k} \pm 2{\Lambda_{L}}^{5/2}\sqrt{\lambda M +m_{Q}}
      =\sum_{k=2}^{N_c}\frac{g_{k}{\rm Tr}{\Phi^{k}_{cl}}}{k} \pm 2 g_{N_c}{\Lambda}^{N_c-\frac{1}{2}}\sqrt{\lambda M +m_{Q}}. 
\label{$9$}
\end{equation}
When we integrated out the fluctuation of the adjoint superfield, we assumed the v.e.v.s of the quarks to be zero. The v.e.v.s of the $SU(2)$ invariant operator, $\bra Q\tilda{Q} \ket$ is order of ${\Lambda_{L}}^{5/2}[\lambda M+m_{Q}]^{-\frac{1}{2}}= \Lambda^{N_c-\frac{1}{2}} g_{N_c} [\lambda M+m_{Q}]^{-\frac{1}{2}}$, so we need ${\it m^{\rm 2}_{{\rm SU(2)}}} \gg \Lambda^{N_c-\frac{1}{2}} g_{N_c} [\lambda M + m_{Q}]^{-\frac{1}{2}}$. It is required that the  v.e.v.s of the $SU(2)$ invariant operators should satisfy the condition like this, also in the case of SQCD with general flavors. 
From $W_{LL}$ we can obtain the equations which relate the quantum moduli to the classical moduli by taking the derivative of $W_{LL}$ with respect to $g_r$,
\begin{equation}
u_{r}={u^{cl}}_{r} \pm 2\Lambda^{N_{c}-\frac{1}{2}}\Big(\frac{g_{N_c}\lambda {\partial {M} \over \partial {g_r}}}{2\sqrt{\lambda M+m_{Q}}}+\sqrt{\lambda M +m_{Q}}\delta_{N_{c},r}\Big).
\label{$11$}
\end{equation}
From the definition of $M$, that is $M=\frac{g_{N_c-1}}{g_{N_c}}$, we get
\begin{equation}
u_{r}={u^{cl}}_{r} \pm 2 \Lambda^{N_{c}-\frac{1}{2}}\Big[\delta_{N_{c},r}(\sqrt{\lambda M + m_Q}-\frac{\lambda M}{2 \sqrt{\lambda M + m_Q}})+\delta_{N_{c}-1,r}\frac{\lambda}{2\sqrt{\lambda M+m_Q}}\Big].
\label{$12$}
\end{equation}
Now using the characteristic equation and the Newton formula, 
\eq
P(x: u)=\bra \det(x-\Phi) \ket =\sum_{k=0}^{N_c} {s}_k x^{N_{c}-k},\ \ \
k s_{k}=-\sum_{j=1}^{k} j s_{k-j} u_j,
\label{$14$}
\end{equation}

we get the relations between classical ${s^{cl}}_r$ and quantum $s_r$,
\begin{eqnarray}
s_{k}&=&s^{cl}_k \ \ \ \  \ k=0,1,2\cdots N_{c}-2 ,\nonumber\\
s_{N_c-1}&=&{s^{cl}}_{N_c-1} \mp \Lambda^{N_{c}-\frac{1}{2}} \frac{\lambda}{\sqrt{\lambda M+m_Q}} ,     \label{a}  \\
s_{N_c}&=&{s^{cl}}_{N_c} \mp \Lambda^{N_{c}-\frac{1}{2}}\frac{\lambda M +2 m_{Q}}{\sqrt{\lambda M + m_Q}}.      \nonumber
\end{eqnarray}
Using (\ref{a}) to rewrite $P(x:u_{cl})=\sum_{k=0}^{N_c} {s}_k^{cl}
x^{N_{c}-k}$ by $s_k$, we get
\begin{equation}
P(x: u_{cl})_{\pm}=P(x: u) \pm \Lambda^{N_{c}-\frac{1}{2}} {(\lambda M+m_Q)}^{-\frac{1}{2}} (\lambda x + \lambda M + 2m_Q).
\label{c}
\end{equation} 
and consider two vacua together by defining
\begin{equation}
\hspace*{-1.0cm}\tilda {P}(x) \equiv P(x: u_{cl})_{+} P(x: u_{cl})_{-} = {P(x: u)}^2 - \Lambda^{2N_{c}-1} {(\lambda M+m_Q)}^{-1} {(\lambda x+\lambda M + 2m_Q)}^2.
\label{$18$}
\end{equation}
Either $P(x: u_{cl})_{+}$ or $P(x: u_{cl})_{-}$ has the double root at $x=M$, so we have
\begin{eqnarray}
{\tilda {P}}(x=M)&=&     P(M: u_{cl})_{+} P(M: u_{cl})_{-}   \nonumber \\
                 &=& {P(M: u)}^2 - \Lambda^{2N_{c}-1} (\lambda M+m_{Q})^{-1} (\lambda M+\lambda M + 2m_{Q})^2    \nonumber\\
                &=&{P(M: u)}^2 - 4\Lambda^{2N_{c}-1} (\lambda M+m_{Q})=0,
\nonumber\\
\left.\frac{d\tilda {P}(x)}{dx}\right|_{x=M} &=& \left. 2P(M:u)\frac{dP(x:u)}{dx}\right|_{x=M}- 4\Lambda^{2N_{c}-1}\lambda=0,
\nonumber
\end{eqnarray}
which show that
\begin{equation}
y^2=P(x:u)^2 - 4\Lambda^{2N_{c}-1}(\lambda x+m_Q)        \nonumber
\end{equation}
is singular at $x=M$. This is the known $N=2$ curve for $SU(N_c)$ with
one flavor \cite{Arg} when the quark masses and the yukawa couplings reduce to
those of $N=2$, that is $\lm_{ij}=\delta_{ij}$ and
$[m,m^{\dagger}]=0$. The authors of \cite{Eli} have derived this
result by the ``integrate-in'' method \cite{Int} for $\Phi$ to get the superpotential with $Q$, $\tilde Q$, and $\Phi$, and by the equations of motion for this potential.

Next consider the case of two flavors. We have to redefine the quarks
in order to make clear the transformation property under the flavor
symmetry, enhanced by the residual $SU(2)$ gauge symmetry as
$Q^{2r-1,\alpha} \equiv Q^{r,\alpha}$, $Q^{2s,\alpha} \equiv
\tilda{Q^{s}}_{\beta} \epsilon^{\alpha \beta}$
\footnote{$\epsilon^{\alpha \beta}$ and $\epsilon_{\alpha \beta}$ are
antisymmetric tensors, $\epsilon^{12}=-\epsilon^{21}=1$ and
$\epsilon_{21}=-\epsilon_{12}=1$.} and we denote $r, s= 1 , 2$ as the
original flavor indices and $i,j = 1 ,2 ,3, 4$ as those of the
enhanced flavor symmetry. $\alpha,\beta$ mean the indices of $SU(2)$
gauge symmetry. Then the $SU(2)$ gauge invariant operators are $V^{i
j} \equiv \epsilon_{\alpha \beta} {Q^i}^{\alpha} {Q^j}^{\beta}$.  As
in the case of one flavor, we obtain the low energy superpotential
with the $SU(2)$ quarks,
\begin{equation}
W_{L} = \sum_{s,r =1,2}({\lambda}_{r,s} M + {m_{Q}}_{r,s})V^{2r-1,2s} + ({\rm Pf}V - {   \Lambda_L}^4)Y+ \sum_{k=2}^{N_c}g_r {u^{cl}}_r,     \nonumber\\
\end{equation}  
where $\Lambda_L$ is the dynamical scale of $SU(2)$ with two flavors and $\Lambda_L^{4}={g_{N_c}}^{2} {\Lambda}^{2N_{c}-2}$ \cite{ter}, \cite{Kut}.  ${\rm Pf}V$ is the 
 Pfaffian, ${\rm Pf}V = \frac{1}{2^{N_c} N_c!} \epsilon_{i_1 i_2 j_1 j_2 \cdots} V^{i_1 i_2}      V^{j_1 j_2} \cdots   $ 
 and $Y$ is a Lagrange multiplier superfield to the constraint for $V$ due to the quantum effect in $SU(2)$ gauge theory with two flavors, ${\rm Pf}V = {\Lambda_L}^4$ \cite{Sei}. Integrating out $V$, we get the superpotential containing the $SU(2)$ gaugino condensation, 
\eq
W_{LL} =\pm  2\Lambda_{L}^{2} [\det(\lambda M + m_Q)]^{\frac{1}{2}} + \sum_{k=2}^{N_c}g_r u^{cl}_r\nonumber\\      = \pm 2 \Lambda^{N_{c}-1} g_{N_c} [\det(\lambda M + m_Q)]^{\frac{1}{2}} + \sum_{k=2}^{N_c}g_r u^{cl}_r,
\label{$22$}
\en
where det means the determinant for the original flavor indices, that is the determinant of $2 \times 2$ matrix. From this we can derive the relation between the quantum moduli and the classical moduli,
\begin{eqnarray}
&&\hspace*{-1.0cm}u_{r} = u^{cl}_{r} \pm  \Lambda^{N_{c}-1} [\det(\lambda M + m_{Q})]^{\frac{1}{2}}
 \Big[ 2 \delta_{r,N_{c}} + \frac{g_{N_c}}{\det(\lambda M + m_{Q})} \frac{\pa M}{\pa g_r} \frac{\pa \det(\lambda M + m_{Q})}{\pa M} \Big],  \label{uuu}\\
&&\hspace*{-1.0cm}\frac{\pa M}{\pa g_r}= - \frac{M}{g_{N_c}} \delta_{r,N_{c}} + \frac{1}{g_{N_c}} \delta_{r,N_{c}-1} .    \label{uu}
\end{eqnarray}
As in (\ref{a}), from (\ref{uuu}), (\ref{uu}) and the Newton formula,
\begin{eqnarray}
s_{k}&=&{s^{cl}}_{k} \ \ \ \ \ k=0,1,2\cdots N_{c}-2  ,\nonumber\\
s_{N_{c}-1}&=&{s^{cl}}_{N_{c}-1} \mp  \Lambda^{N_{c}-1}[\det(\lambda M + m_{Q})]^{-\frac{1}{2}} {\partial{\det( \lambda M + m_{Q})}\over \partial {M}} ,\label{ss}\\
s_{N_c}&=&{s^{cl}}_{N_c} \mp  \Lambda^{N_{c}-1} \Big(2 [\det(\lambda M + m_{Q})]^{\frac{1}{2}} - \frac{M}{[\det( \lambda M + m_{Q})]^{\frac{1}{2}}} 
      \times {\partial{\det( \lambda M + m_{Q})} \over \partial {M}}\Big)
\nonumber
\end{eqnarray}
are obtained. By using these results we define $\tilda {P} (x)$ as in (\ref{$18$}),
\begin{eqnarray}
{\tilda {P}}(x)&=&P(x: u_{cl})_{+} P(x: u_{cl})_{-} \nonumber\\
     &=& {P(x: u)}^2 - \Lambda^{2 N_{c}-2} \Big(x [\det(\lambda M + m_{Q})]^{-\frac{1}{2}} {\partial{\det( \lambda M + m_{Q})} \over \partial {M}}\nonumber \\
     &+&2[\det(\lambda M + m_{Q})]^{\frac{1}{2}} - \frac{M}{[\det( \lambda M + m_{Q})]^{\frac{1}{2}}} {\partial{\det( \lambda M + m_Q)} 
   \over \partial {M}}\Big)^2.  \label{tt}
\end{eqnarray}
Substituting $x=M$, and by the condition of the double root we get
\begin{eqnarray}
&&\hspace*{-1.0cm}{\tilda {P}}(x=M)=P(M: u^{cl})_{+} P(M: u^{cl})_{-} ={P(M: u)}^2 - 4 \Lambda^   {2N_{c}-2} \det(\lambda M + m_{Q}) = 0 , \label{$266$}  \\
&&\hspace*{-1.0cm}\left. \frac{d \tilda {P}(x)}{dx} \right|_{x=M} = 2 P(M:u) \left.\frac{dP(x:u)}{dx}\right|_{x=M} -  4 \Lambda^{2N_{c} -2} {\partial{\det(\lambda M + m_{Q})} \over \partial {M}} =0.    
\label{$26$}   
\end{eqnarray}
The above equations mean that the curve described by
\begin{equation}
y^2=P(x:u)^2 - 4\Lambda^{2N_{c}-2}\det(\lambda x+m_{Q}) \nonumber
\end{equation}
is singular at $x=M$. This is the $N=2$ curve for $SU(N_c)$ with two flavors \cite{Arg} in the $N=2$ limits of the couplings. We have neglected $W_{\Delta}$ in the steps above as before.

We now consider the general theory with $N_f$ flavors. ($2<N_f < N_c$)
\begin{equation}
W_{L} = \sum_{s,r =1}^{N_f}(\lambda_{r,s} M + {m_{Q}}_{r,s})V^{2r-1,2s} + ({\rm Pf}V)^{\frac{1}{N_f - 2}}(2-N_f) {\Lambda_L}^{\frac{N_f-6}{N_f-2}} + \sum_{k=2}^{N_c}g_r {u^{cl}}_r ,    \nonumber
\end{equation}
where we redefine the quarks as before, 
$Q^{2r-1,\alpha}\equiv Q^{r,\alpha}$ ,
$Q^{2r,\alpha} \equiv {\tilda{Q^{r}}}_{\beta}\epsilon^{\alpha \beta}$ ,
$V^{i j} \equiv \epsilon_{\alpha \beta} {Q^i}^{\alpha} {Q^j}^{\beta},$ 
and we denote the relation between the original dynamical scale, $\Lambda$ and that of the low energy $SU(2)$ with $N_f$ flavors, $\Lambda_L$ as ${\Lambda_L}^{6-N_f}={g_{N_c}}^{2} {\Lambda}^{2N_{c}-N_f}$ \cite{ter}, \cite{Kut}.  
Here, the second term in $W_L$ is the effective superpotential determined by the symmetry in $SU(2)$ gauge theory with $N_f$ flavors, \cite{Sei} and $r, s= 1 , 2\cdots N_f $ are the indices of the original flavor symmetry while $ i,j = 1 ,2 ,3 \cdots 2N_f$ are the indices of the flavor symmetry, enhanced by the residual $SU(2)$ gauge symmetry. 
From the equations of motion for $V$, we get $\bra {\rm Pf}V \ket = g_{N_c}^{N_f} [\det(\lambda M + m_Q)]^{\frac{N_f-2}{2}} \Lambda^{N_f(N_{c} - N_{f}/2)}$ and obtain the superpotential with the $SU(2)$ gaugino condensation,
\begin{equation}
W_{LL}=\sum_{k=2}^{N_c}g_r u^{cl}_r \pm 2 [\det(\lambda M + m_Q)]^{\frac{1}{2}} g_{N_c} \Lambda^{N_{c} - N_{f}/2},
\label{$30$}
\end{equation}
where det means the determinant for the $N_f \times N_f$ matrix of the original flavor symmetry. We assume that $W_{\Delta}$ is zero. It is important to note that in the case of $N_{f}>3$, the couplings must be far away from the points where $\bra {\rm Pf}V \ket=0$ is satisfied, because the dual quarks contribute there. Of course, in addition to this condition,  $\bra V \ket$ must be much smaller than ${\it m^{\rm 2}_{{\rm SU(2)}}}$ as in the case of one flavor. 


(\ref{$30$}) is the same as (\ref{$22$}) if we replace the determinant in (\ref{$22$}) by that of the $N_f \times N_f$ flavor matrix and replace $\Lambda^{N_{c}-1}$ in (\ref{$22$}) by $\Lambda^{N_{c}-N_f/2}$. So we can conclude from (\ref{$30$}) that the curve,   
\begin{equation}
y^{2}=P(x:u)^2-4 \Lambda^{2N_{c}-N_{f}} \det(\lambda x + m_{Q})
\end{equation}
becomes singular at $x=M$, in the same way as (\ref{uuu}), (\ref{uu}), (\ref{ss}),  (\ref{tt}), (\ref{$266$}) and (\ref{$26$}) in the case of two flavors. This curve corresponds to the $N=2$ curve for $SU(N_c)$ $N_f$ flavor SQCD \cite{Arg} in the $N=2$ limits of the couplings. 

If we scale $\tilda {\Lambda}^{2N_{f}}= \det(m_{Q}) \Lambda^{2N_{c}-N_{f}}$ with large $m_Q$, then in (\ref{$30$}) we get the same superpotential of \cite{Eli} by which the authors have derived the $N=2$ SYM curve \cite{Kle}, \cite{Far}. That is, it corresponds to integrating all the quarks first and considering remaining $N=2$ pure gauge theory by perturbing with $W_{tree}$  without quark mass terms and  yukawa terms.

We have considered the case of $N_f < N_c$ until now. For $ N_f \geq N_c $, if we add the baryon source terms,
\begin{equation}
b_{i_{1}i_{2}\cdots i_{N_{c}}}B^{i_{1}i_{2}\cdots i_{N_{c}}} + \tilda {b}^{i_{1}i_{2}\cdots i_{N_{c}}}\tilda {B}_{i_{1}i_{2}\cdots i_{N_{c}}}
\label{$33$}
\end{equation}
to $W_{tree}$, $W_{LL}$ is the same as in $N_{f} < N_{c}$ in our
method because these terms do not seem to contribute when $M_a$
are very large. Then the curves are the same as in $N_{f} < N_{c}$ and
correspond with the results of \cite{Arg} in the case of $N_f <
2N_c-1$. For $N_f=2N_c-1$, $2N_c$, we need more modifications of this
method to be consistent with the $N=2$ results of \cite{Arg}. Compared with
\cite{Han}, our results are the same as in $N_{f} < N_{c}$. But in
$N_{f} \geq N_{c}$, their results are different from ours. It may be due
to the constant shifts of the renomarization for the moduli because the moduli are composite.

In summary, we have derived the conditions for the singularities of the
curves with general yukawa couplings and quark masses in the case of
$SU(N_c)$ with flavors by using the $N=1$ effective superpotential
with the gaugino condensation. In the $N=2$ limits of the couplings,
these curves are the same as the known ones of $N=2$ SQCD. These
results are consistent with condensation of massless solitons
that we expect from $N=2$ theory.




\vskip5mm
The author would like to thank K. Izawa and T. Yoneya for useful
discussions. It is pleasure to appreciate J. Furukawa, K. Itakura, S. Terashima, K. Tobe and S.-K. Yang, for critical comments.




\begin{thebibliography}{99} 
\bibitem{Wit} N. Seiberg and E. Witten, hep-th/9407087,
              Nucl. Phys. {\bf B426} (1994) 19.

\bibitem{aps}  P. C. Argyres, M. R. Plesser and N. Seiberg, hep-th/9603042,
               Nucl. Phys. {\bf B471} (1996) 159.  

\bibitem{Eli}S. Elitzur, A. Forge, A. Giveon, K. Intriligator and 
             E. Rabinovici, hep-th/9603051, Phys. Lett. 
              {\bf B379} (1996) 121.

\bibitem{Kle} A. Klemm, W. Lerche, S. Yankielowicz and S. Theisen, 
              hep-th/9411048, Phys. Lett. {\bf 344} (1995) 169.

\bibitem{Far} P. C. Argyres, and A. E. Faraggi, hep-th/9411057, 
              Phys. Rev. Lett. {\bf 74} (1995) 3931.

\bibitem{ter} S. Terashima and S-K. Yang, hep-th/9607151.

\bibitem{Gid} K. Landsteiner, J. M. Pierre and S. B. Gidding, hep-th/9609059.

\bibitem{Int} K. Intriligator, hep-th/9407106, 
              Phys. Lett. {\bf B336} (1994)  409.

\bibitem{Arg} P. C. Argyres, M. R. Plesser and A. D. Shapere, hep-th/9505100, 
              Phys. Rev. Lett. {\bf 75} (1995) 1699.

\bibitem{Kut} D. Kutasov and A. Schwimmer and N. Seiberg, hep-th/95010222,    
              Nucl. Phys. {\bf B459} (1996) 455. 

\bibitem{Sei} N. Seiberg, hep-th/9402044, Phys. Rev. {\bf D49} (1994) 6857.  

\bibitem{Han} A. Hanany and Y.Oz, hep-th/9505075,
                 Nucl. Phys. {\bf B452} (1995) 283. 
              

 
\end{thebibliography}
\end{document}